\documentclass[letterpaper]{article}
\usepackage[]{nexteditprediction} 
\usepackage{times} 
\usepackage{helvet} 
\usepackage{courier} 
\usepackage[hyphens]{url} 
\usepackage{graphicx}
\usepackage{natbib}
\usepackage{caption}
\usepackage{algorithm}
\usepackage{algorithmic}

\usepackage{newfloat}
\usepackage{listings}
\DeclareCaptionStyle{ruled}{labelfont=normalfont,labelsep=colon,strut=off}
\floatstyle{ruled}
\newfloat{listing}{tb}{lst}{}
\floatname{listing}{Listing}

\title{Next Edit Prediction: Learning to Predict Code Edits from Context and Interaction History}
\author {
    Ruofan Lu\textsuperscript{\rm 1},
    Yintong Huo\textsuperscript{\rm 2},
    Meng Zhang\textsuperscript{\rm 3},
    Yichen Li\textsuperscript{\rm 1}\thanks{Yichen Li is the corresponding author.},
    Michael R. Lyu\textsuperscript{\rm 1}
}
\affiliations {
    \textsuperscript{\rm 1}The Chinese University of Hong Kong\\
    \textsuperscript{\rm 2}Singapore Management University\\
    \textsuperscript{\rm 3}TabbyML\\
    \{rflu, ycli21, lyu\}@cse.cuhk.edu.hk, ythuo@smu.edu.sg, meng@tabbyml.com
}

\usepackage{bibentry}

\usepackage{float}
\usepackage{amsmath}
\usepackage{graphicx}
\usepackage{booktabs}
\usepackage{pgfplots}
\usepackage{subcaption}
\usepackage{multirow}
\usepackage{makecell}
\usepackage{enumitem}
\usepackage{hyperref}
\pgfplotsset{compat=newest}
\usepgfplotslibrary{statistics}

\begin{document}

\maketitle

\begin{abstract}
The rapid advancement of large language models (LLMs) has led to the widespread adoption of AI-powered coding assistants integrated into a development environment. On one hand, low-latency code completion offers completion suggestions but is fundamentally constrained to the cursor's current position. On the other hand, chat-based editing can perform complex modifications, yet forces developers to stop their work, describe the intent in natural language, which causes a context-switch away from the code. This creates a suboptimal user experience, as neither paradigm proactively predicts the developer's next edit in a sequence of related edits.
To bridge this gap and provide the seamless code edit suggestion, we introduce the task of \textit{Next Edit Prediction}, a novel task designed to infer developer intent from recent interaction history to predict both the location and content of the subsequent edit.
Specifically, we curate a high-quality supervised fine-tuning dataset and an evaluation benchmark for the \textit{Next Edit Prediction} task. Then, we conduct supervised fine-tuning on a series of models and performed a comprehensive evaluation of both the fine-tuned models and other baseline models, yielding several novel findings. 
This work lays the foundation for a new interaction paradigm that proactively collaborate with developers by anticipating their following action, rather than merely reacting to explicit instructions. The code is available at \href{https://github.com/lurf21/NextEditPrediction}{https://github.com/lurf21/NextEditPrediction}.
\end{abstract}

\section{Introduction}
\label{sec:intro}

In recent years, the advancement of large language models (LLMs) has led to the emergence of numerous AI coding assistants with millions of users. These tools provide features such as code completion \cite{husein2024large, izadi2024language} and code editing \cite{chakraborty2020codit, zhang2022coditt5, cassano2023can}, substantially enhancing the efficiency of development workflows and consequently achieving widespread adoption. However, they primarily operate within two distinct interaction paradigms, each of which presents a significant barrier to the seamless user experience.

The first and most common paradigm is low-latency code completion with Fill-in-the-Middle (FIM) techniques~\cite{FIM}, where models generate suggestions based on code snippets before and after the cursor position. While effective for local completions, it is fundamentally constrained to the cursor's position. The second paradigm employs the chat-based interface where users describe editing intentions through natural language~\cite{cassano2023can}, forcing developers to be distracted from their creative flow.

The limitations of chat-based interaction are particularly evident in modern agentic code assistants. These powerful agents, which adopt the chat paradigm, are capable of tackling complex tasks, such as building entire web applications from scratch. However, their practical application in code editing is hindered by several drawbacks. Firstly, they require users to explicitly describe their intentions for every complex step. Secondly, their reliance on powerful and slow backbone models often leads to high response latency, which is unsuitable for the near real-time feedback. Lastly, the process of reviewing and managing multiple, model-generated changes does not align with typical, granular editing workflows. These issues highlight a critical need for a new interaction mode that merges the proactive nature of code completion with the semantic power of instruction-based editing from the past edit sequences.

\begin{figure*}[htbp]
    \centering
    \includegraphics[width=0.8\textwidth]{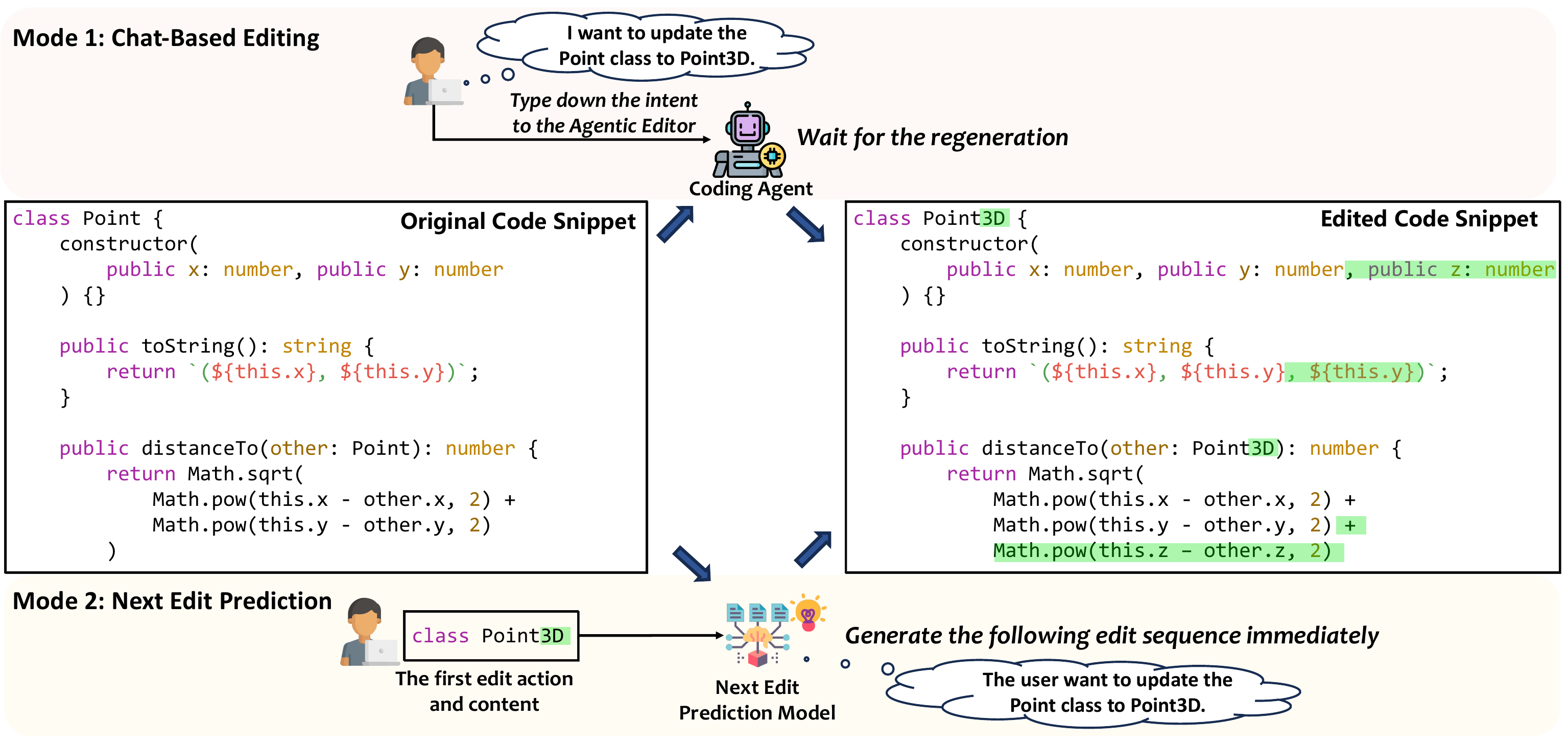}
    \caption{Comparison of AI coding assistant workflows based on Chat-Based Editing and Next Edit Prediction.}
    \label{fig:comparison}
\end{figure*}

To enable this truly fluid code editing experience, we propose \textit{Next Edit Prediction} as a novel task and interaction paradigm that requires models to predict both the location and content of subsequent edit by analyzing evolving code context and edit history. As illustrated in Figure~\ref{fig:comparison}, Unlike traditional chat-based editing, our approach proactively infers developer intentions through dynamic analysis of code edit patterns without interrupting the current flow of developer. The task takes the proceeding code edit sequence as input, including both the current code context and historical edit actions, to generate the potential next edit. This temporal reasoning enables prediction of next-step changes like API updates, type/object changes, and identifier renaming~\cite{CodeEditor}, suggesting context-aware edits that extend developer workflows while handling complex jump-location modifications beyond single-point completions without natural language instruction.

To support this task, we create a dataset and a benchmark through systematic engineering. We build: (\textit{i}) \textbf{A context-aware dataset} derived from CommitPackFT~\cite{Muennighoff24octopack}, implementing multi-criteria filtering that eliminates 72.8\% of incoherent edits via in-context binary classification, resulting in 3,211 high-fidelity samples (2M tokens) across seven languages; (\textit{ii}) \textbf{A human-labelled benchmark} comprising 210 manually verified commits from exclusive top-100 starred GitHub repositories, built through file-type pre-screening, training-aligned rule constraints, and interactive diff visualization for human validation. 

In summary, our contributions are as follows: (\textit{i}) We are the first to introduce the task of \textit{Next Edit Prediction}.
(\textit{ii}) We constructed a dataset and a benchmark for training and evaluating the \textit{Next Edit Prediction} task.
(\textit{iii}) Our experiments with fine-tuned open-source models demonstrate the potential of smaller models for this task, offering valuable insights for the development of future code editing paradigm.


\section{Dataset and Benchmark}
\begin{figure*}[htbp]
    \centering
    \includegraphics[width=0.8\textwidth]{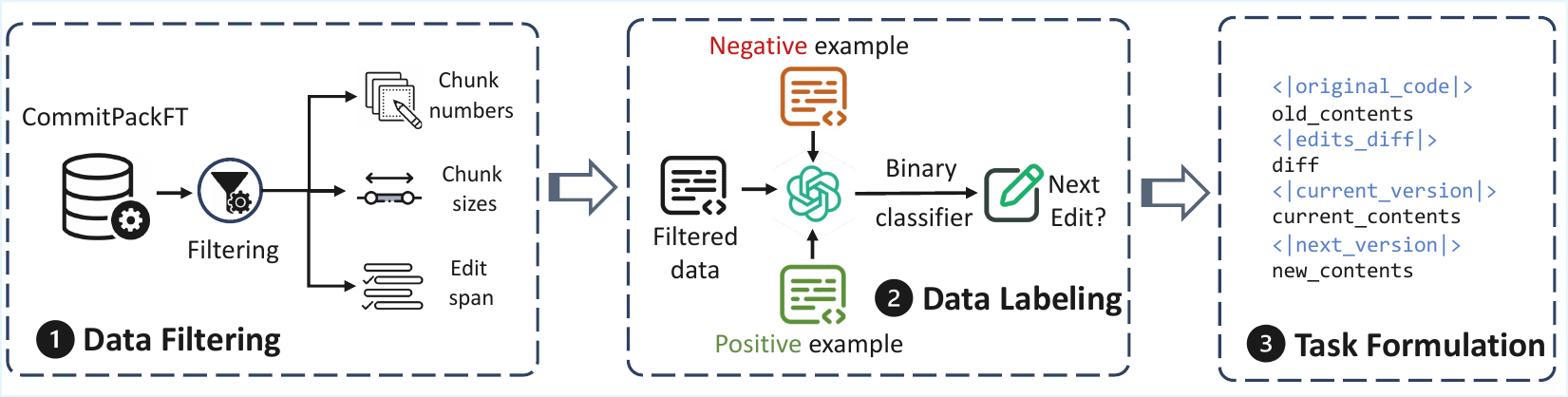}
    \caption{Dataset curation pipeline.}
    \label{fig:curation}
\end{figure*}

\label{sec:dataset_benchmark}
For the task of \textit{Next Edit Prediction}, the objective is to identify sequences of edits where prior changes provide meaningful context, enabling the profiling of the next edit intent for predicting subsequent modifications.

Code commits serve as an ideal foundation for the \textit{Next Edit Prediction} task. Each commit represents a logical unit of code evolution, consisting of chronologically ordered edit chunks. These commit-based edit sequences inherently capture real-world patterns of developer reasoning, where prior modifications establish context for subsequent changes.

\subsection{Dataset Curation}
\label{sec:dataset}
CommitPack~\cite{muennighoff2024octopackinstructiontuningcode} is a dataset constructed from commit data in GitHub repositories. We selected it as the data source and built a dataset for supervised fine-tuning through three steps: Data Filtering, Data Labeling, and Task Formulation.

\subsubsection{Data Filtering.}
The CommitPack dataset is available in two versions: the full version comprises 4 terabytes of data and covers 350 programming languages. Based on this, a stricter filtering process was applied to reduce the dataset to 2 gigabytes, encompassing 277 programming languages—this filtered version is referred to as CommitPackFT. In this work, we focus primarily on the seven most commonly used programming languages (i.e., Python, Java, Go, C, C++, JavaScript, and TypeScript). We use the data corresponding to these seven languages in CommitPackFT as the source for data filtering.

To isolate high-quality training candidates, we applied a strict set of filtering criteria to the raw commit data. We define a contiguous sequence of edited lines as an \textit{edit chunk}. Our criteria are as follows:
\begin{enumerate}[itemsep=2pt, parsep=0pt]
    \item \textbf{Multiple Edit Chunks:} A commit must contain at least two edit chunks. This ensures there is a history of at least one edit to serve as the resource for a subsequent edit.
    \item \textbf{Bounded Chunk Length:} Each edit chunk must not exceed five lines. Overly long, complex changes fall outside this scope and are considered to be different types of interaction.
    \item \textbf{Limited Edit Scope:} The total distance between the first edit chunk and the last line of the last edit chunk within a commit must not exceed \textit{80} lines. This prevents the context from becoming excessively large and unfocused.
    \item \textbf{Additive Edits Only:} To simplify the initial task formulation, we select only commits that consist exclusively of additive edits, excluding those with deletions.
\end{enumerate}


\subsubsection{Data Labeling.}
Raw filtering is insufficient, as many commits contain semantically unrelated edits (e.g., fixing a typo and separately refactoring a function). To address this, we employ a labeling stage to identify semantically coherent edit sequences. We utilize the in-context learning~\cite{dong2022survey} strategy for GPT-4o mini~\cite{gpt4o} with labelled examples to perform binary classification. The model is prompted to analyze a sequence of edits and determine if the final edit is a logical continuation of the preceding ones.

\begin{figure}[H]
  \centering
  \begin{subfigure}[t]{0.95\columnwidth}
    \centering
    \includegraphics[width=\textwidth]{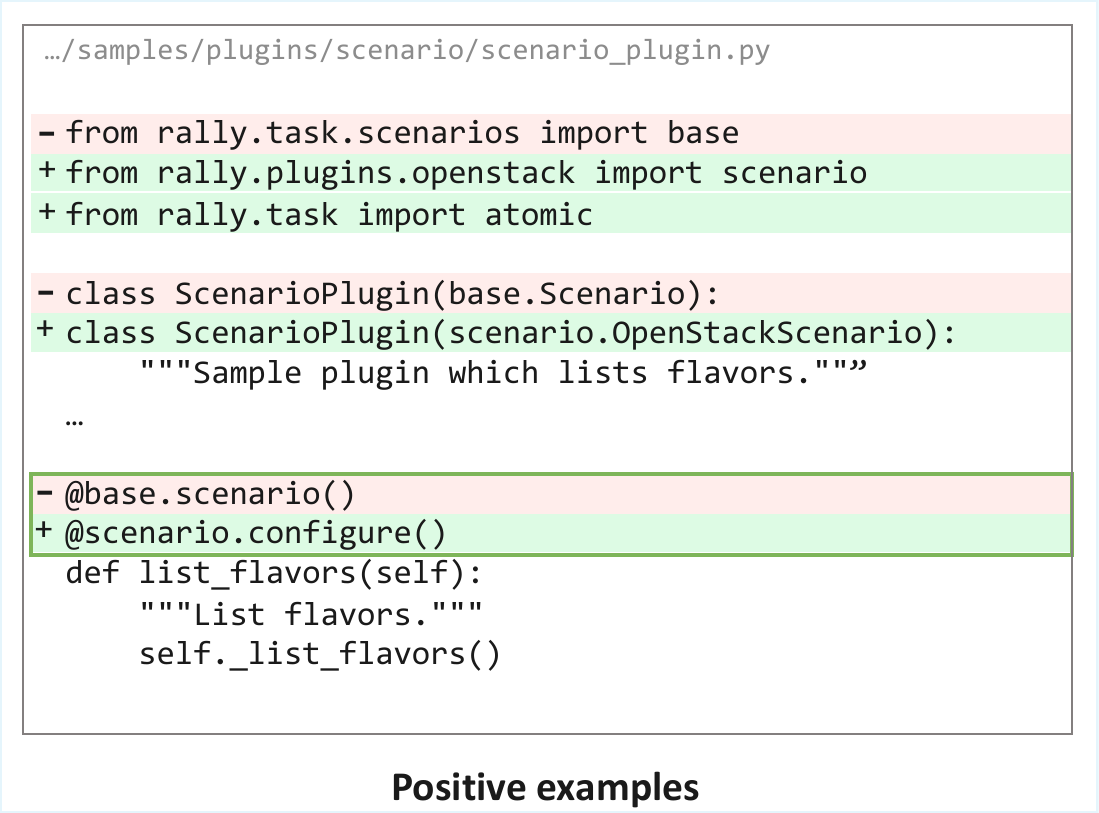}
    \caption{}
    \label{fig:img1}
  \end{subfigure}
  \hfill
  \begin{subfigure}[t]{0.95\columnwidth}
    \centering
    \includegraphics[width=\textwidth]{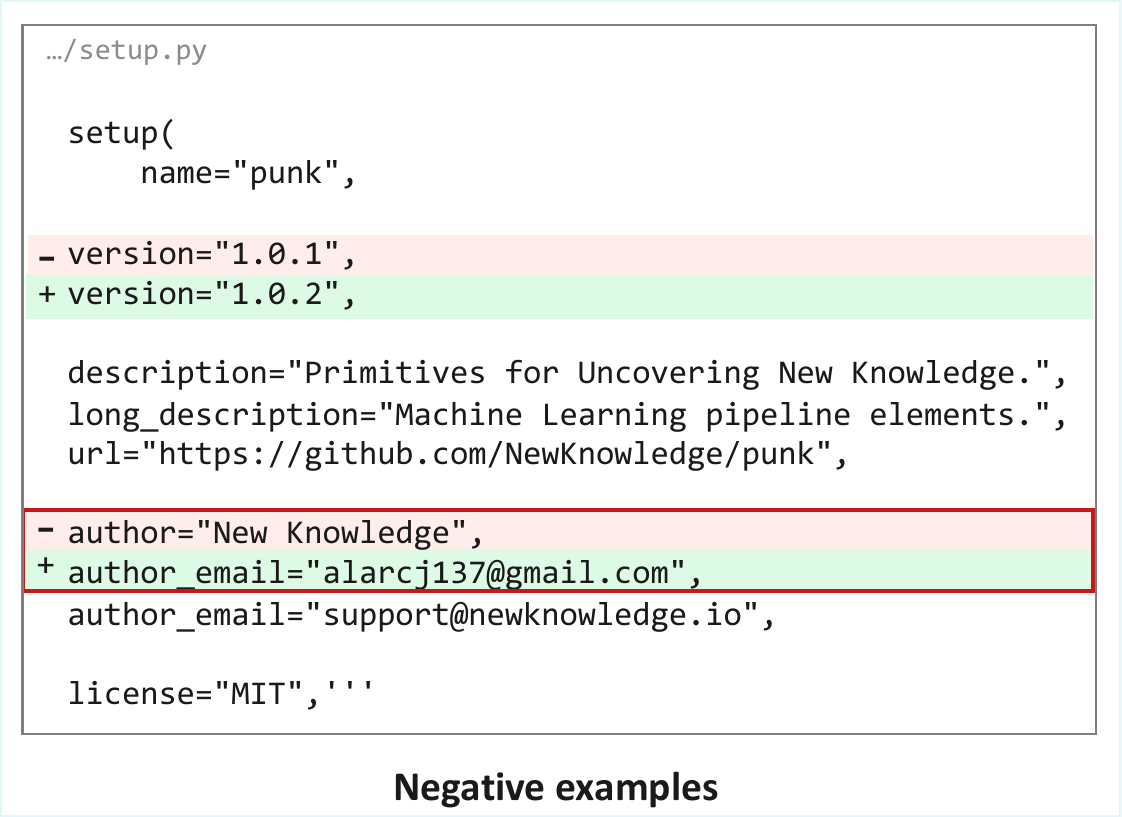}
    \caption{}
    \label{fig:img2}
  \end{subfigure}
  \caption{The positive and negative examples used for data labeling.}
  \label{fig:examples}
\end{figure}

Figure~\ref{fig:examples} illustrates this process by showing the exact types of examples provided as one of the given demonstrations to the labeling model.

The positive example, shown in Figure~\ref{fig:img1}, demonstrates a sequence of semantically linked edits. Here, multiple changes collaborate to refactor a class from one framework to another, which includes updating the import-use chain (from \texttt{base.Scenario} to \texttt{scenario.OpenStackScenario}) and modifying the decorator syntax (from \texttt{@base.scenario()} to \texttt{@scenario.configure()}). This is a high-quality sample because each edit provides crucial context for predicting the next, collectively implementing a consistent framework migration.

Conversely, the negative example in Figure~\ref{fig:img2} shows unrelated edits within a single commit. While the first edit increments a version number (\texttt{"1.0.1"} to \texttt{"1.0.2"}), the second changes an author's email address. These modifications serve distinct purposes with no predictive relationship, making the sequence an inappropriate training sample. By providing these clear, contrasting examples in the in-context learning strategy, we guide the model to effectively distinguish between meaningful, predictive edit sequences and coincidental, unrelated ones.

\subsubsection{Task Formulation.}
After completing data filtering and labeling, we format the data for supervised fine-tuning. Similar to the Fill-in-the-Middle paradigm, we introduce four new tokens: \texttt{<|original\_code|>}, \texttt{<|edits\_diff|>}, \texttt{<|current\_version|>}, and \texttt{<|next\_version|>}, to assist in constructing each sample. The final format of the sample is illustrated as follows:
\[
\begin{array}{l}
\texttt{<|original\_code|>}\\
\texttt{old\_contents}\\
\texttt{<|edits\_diff|>}\\
\texttt{diff}\\
\texttt{<|current\_version|>}\\
\texttt{current\_contents}\\
\texttt{<|next\_version|>}\\
\texttt{new\_contents}
\end{array}
\]
\texttt{old\_contents} and \texttt{diff} represent the edit history, \texttt{current\_contents} refers to the current code context, and \texttt{new\_contents} denotes the code segment that the model is expected to predict based on the context and the edit history as a possible next edit.

\begin{table*}[htbp]
\centering
\small
\begin{tabular}{l|ccccc}
\toprule
\textbf{Languages} & \textbf{Positive} & \textbf{Filtered} & \textbf{Raw} & \textbf{Positive/Filtered Ratio} & \textbf{Average Length} \\ \midrule
Python & 1,056 & 1,330 & 56,025 & 79.40\% & 684 \\
Java & 655 & 709 & 20,635 & 92.38\% & 751 \\
Go & 192 & 227 & 5,004 & 84.58\% & 737 \\
C & 220 & 314 & 8,506 & 70.06\% & 731 \\
C++ & 130 & 180 & 4,992 & 72.22\% & 720 \\
JavaScript & 832 & 1,056 & 52,989 & 78.79\% & 709 \\
TypeScript & 126 & 139 & 5,868 & 90.65\% & 739 \\ \midrule
All & 3,211 & 3,955 & 154,019 & 81.19\% & 714 \\
\bottomrule
\end{tabular}
\caption{Data statistics of training dataset.}
\label{table:data}
\end{table*}
\subsection{Dataset Statistics}
Our dataset contains a total of 3,211 samples, comprising approximately 2M tokens and covering seven programming languages. We present a statistical analysis of the dataset in Table \ref {table:data}. The columns Positive, Filtered, and Raw represent the number of positive-labeled samples in the filtered data, the total number of filtered samples, and the number of initial samples in CommitPackFT, respectively. Additionally, we report the average tokenized length of each sample for each programming language in the dataset, computed using the tokenizer of our fine-tuned Qwen2.5-Coder-7B model.

\subsection{Benchmark Construction}
To ensure a robust and unbiased evaluation of model performance, we constructed a separate benchmark from repositories guaranteed to be unseen during training.

\subsubsection{Repository selection and commit crawling.}
We began by retrieving the top 100 repositories with the highest number of stars for each of the seven programming languages considered. This popularity-based selection ensures that the repositories are not only widely recognized by the developer community but also likely to represent idiomatic and widely adopted coding practices within each language.

Following the initial retrieval, we curated a refined subset of repositories from each language-specific list to serve as benchmark data sources for subsequent analysis. This selection was guided by a multi-faceted set of criteria: (\textit{i}) the total number of commits, used as a proxy for project maturity and development history; (\textit{ii}) the nature of the repository content, prioritizing libraries, frameworks, and applications over tutorials or experimental projects; and (\textit{iii}) the recency and frequency of maintenance activity, as repositories with ongoing development and regular updates are more likely to reflect current best practices and language features. 

To avoid the potential risk of data leakage~\cite{huang2024your}, we implemented a stringent filtering process to ensure that the repositories used in our benchmarking experiments were completely disjoint from those included in our tuning dataset. To enforce this separation, we cross-referenced repository identifiers and metadata to systematically exclude any overlapping or related projects.

Once the evaluation repositories were confirmed to be separate from tuning set, we proceeded to collect commit-level data from each of them. For consistency and reproducibility, we adopted the crawling strategy employed in the CommitPack dataset pipeline. 

\subsubsection{Rule-based filtering and processing.}
The filtering rules are consistent with those used during the dataset construction. However, an additional pre-filtering step based on file types is applied. This is necessary because the crawled commits contain diffs of various file types, such as configuration files and Markdown documents, identified by their file extensions. After filtering, the data is processed using the same task formulation method as in the dataset, ensuring consistency between the training and evaluation procedures.

\subsubsection{Manual selection.}
While our automated labeling process is effective for creating the tuning set, a benchmark for evaluating model performance demands an even higher standard of precision and relevance.

To guarantee the relevance and quality of every sample, we developed a web-based interface for previewing the filtered data. This interface displays the results in a unified diff format, enabling efficient browsing and selection of commits. Ultimately, we manually selected 30 commits for each programming language, resulting in a total of 210 samples for the benchmark. The selection process guarantees a significantly high quality benchmark, which forms the foundation for evaluating model performance in real-world code editing tasks.

\begin{table*}[htbp]
\centering
\small
\begin{tabular}{l|cccc}
\toprule
\textbf{Model} & \textbf{Exact Match} & \textbf{Partial Match} & \textbf{Position Match} & \textbf{LLM-as-a-Judge} \\
\midrule

\multicolumn{5}{l}{\textit{\textbf{Closed-source Models}}} \\ \midrule
GPT-4o & 23.33 & 39.05 & 57.14 & 42.86 \\
GPT-4o mini & 27.14 & 35.71 & 46.67 & 37.62 \\
GPT-4.1 & 35.71 & 45.71 & 65.71 & 53.81 \\
GPT-4.1 mini & 40.48 & 47.62 & 62.38 & 50.95 \\
GPT-4.1 nano & 2.86 & 19.05 & 24.76 & 23.81 \\
\midrule
Claude 4 Opus & 45.71 & 67.62 & 80.48 & 67.62 \\
Claude 4 Sonnet & 46.67 & 68.57 & 80.00 & 65.24 \\
Claude 3.7 Sonnet & 33.81 & 59.52 & 75.71 & 61.90 \\
Claude 3.5 Sonnet & 42.86 & 58.10 & 73.33 & 59.52 \\
Claude 3.5 Haiku & 17.14 & 43.33 & 63.81 & 44.76 \\
\midrule
Gemini 2.5 Pro & 46.67 & 62.38 & 76.19 & 68.57 \\
Gemini 2.5 Flash & 43.33 & 61.90 & 77.14 & 61.90 \\
\midrule

\multicolumn{5}{l}{\textit{\textbf{Open-source Models}}} \\ \midrule
DeepSeek-V3 (671B-A37B) & 41.90 & 55.24 & 73.81 & 61.43 \\
DeepSeek-R1 (671B-A37B) & 40.48 & 58.10 & 74.29 & 61.90 \\
\bottomrule
\end{tabular}

\caption{Evaluation results of baseline models.}
\label{table:baseline}
\end{table*}
\section{Experiments}

\subsection{Experimental Settings}
\label{sec:experimental_settings}
\subsubsection{Baselines.}
We compare our fine-tuned models with the following competitive baselines. Closed-Source Models: OpenAI models, Claude models, and Gemini models, specific model names are shown in Table \ref{table:baseline}. Open-Source Models: DeepSeek-R1\cite{deepseekai2025deepseekr1incentivizingreasoningcapability}, DeepSeek-V3\cite{deepseekai2025deepseekv3technicalreport}.

\subsubsection{Supervised fine-tuning.}
We adopt Qwen2.5-Coder-\{3B, 7B, 14B, 32B\}\cite{hui2024qwen25codertechnicalreport}, CodeGemma-\{2B, 7B\}\cite{codegemmateam2024codegemmaopencodemodels}, and Codestral-22B-v0.1\cite{Codestra19:online} as the base models and fine-tune all these models for 2 epochs using a single NVIDIA H100 GPU. For optimal memory efficiency and training effectiveness, we adopt the LoRA\cite{hu2021loralowrankadaptationlarge} technique for fine-tuning. Since we add new tokens into the tokenizer, namely \texttt{<|original\_code|>}, \texttt{<|edits\_diff|>}, \texttt{<|current\_version|>}, and \texttt{<|next\_version|>}, we include the parameters of both the \texttt{lm\_head} and \texttt{embed\_tokens} in the \texttt{target\_modules} for training.
We set the initial learning rate and embedding learning rate at
1e-4 and 1e-5, respectively. The learning
rate is adjusted using a warmup of 5 steps and then decayed following a linear scheduler. We use AdamW optimizer\cite{loshchilov2019decoupledweightdecayregularization} and choose a batch size of 4 with a sequence length of 2,048.

\subsubsection{Evaluation Protocol.}

We adopted a consistent protocol for all evaluations to ensure fairness. For fine-tuned models, the evaluation input format precisely matches the training format to prevent distributional shift. Model inference was performed using the Offline Inference mode of vLLM~\cite{kwon2023efficient} with greedy decoding (temperature = 0) for deterministic outputs. For baseline models not fine-tuned on our task, we used a the one-shot prompting strategy that demonstrates the task format.

\begin{table*}[htbp]
\centering
\small
\begin{tabular}{lc|cccccccc}
\toprule
\textbf{Model} & \textbf{Size} & \textbf{Exact Match} & \textbf{Partial Match} & \textbf{Position Match} & \textbf{LLM-as-a-Judge} \\
\midrule
\multirow{4}{*}{\shortstack[l]{Qwen2.5-Coder}}
& 3B & 33.33 & 40.95 & 58.57 & 42.38 \\
& 7B & 37.62 & 44.29 & 64.76 & 49.05 \\
& 14B & 50.48 & 52.86 & 75.71 & 58.57 \\
& 32B & 51.43 & 52.38 & 73.81 & 60.95 \\
\midrule
\multirow{2}{*}{\shortstack[l]{CodeGemma}}
& 2B & 26.19 & 31.43 & 46.67 & 33.81 \\
& 7B & 43.81 & 49.52 & 62.86 & 50.95 \\
\midrule
Codestral & 22B & 25.24 & 39.05 & 49.05 & 32.86 \\
\bottomrule
\end{tabular}
\caption{Evaluation results of fine-tuned base models. Due to the limited instruction-following capabilities of these base models, using one-shot prompting for evaluation tends to produce a large number of repetitive sequences~\cite{mahaut2025repetitionsalikedistinctmechanisms}. As a result, the pre-fine-tuning results are not reported.}
\label{table:sft_base}
\end{table*}
\subsection{Evaluation Metrics}
\label{sec:evaluation_metrics}
The benchmark is constructed from edit sequences found in large-scale, real-world repositories to ensure the task reflects real development patterns. However, these repositories typically involve complex dependencies and intricate build configurations, making the execution of code snippet difficult.

To address this limitation, we propose four metrics to  comprehensively evaluate model performance from different angles: \textbf{Exact Match}, \textbf{Partial Match}, \textbf{Position Match}, and \textbf{LLM-as-a-Judge}. Among the first three metrics, the level of strictness in evaluating the model’s predictions follows the order: 
\[
\text{Exact Match} > \text{Partial Match} > \text{Position Match}.
\]
We provide detailed definitions of each metric below.

\subsubsection{Exact Match.}
In this benchmark, the most important fields in each data instance are \texttt{old\_contents}, \texttt{current\_contents}, and \texttt{new\_contents}. The task of the model is to predict \texttt{new\_contents} based on \texttt{old\_contents}, \texttt{current\_contents}, and the unified diff generated from the latter two. We define an \textit{Exact Match} as a case where the model's output is identical to \texttt{new\_contents}.

\subsubsection{Partial Match.}
The construction method of the benchmark determines that the \texttt{current\_contents} and \texttt{new\_contents} differ by only one edit chunk. However, considering that the model may sometimes generate multiple new edits at once, the Exact Match metric is no longer applicable in such cases. In this scenario, we divide the unified diff results of the model's output and \texttt{current\_contents} into edit chunks and store them in a list. If the edit chunk that differs between \texttt{current\_contents} and \texttt{new\_contents} is found in the list, it is considered to satisfy the \textit{Partial Match}.

\subsubsection{Position Match.}
\textit{Position Match} focuses on evaluating the model's ability to predict the next editing location, rather than assessing whether the content generated at that position matches the ground truth. The calculation of this metric is similar to that of \textit{ partial match}, with the key difference being that lines within the edit chunk that have been edited (i.e. those starting with \texttt{+} or \texttt{-}) are removed before evaluation. This metric is crucial for correctly profiling the user edit intent, as it verifies whether a predicted code edit targets the same position as the ground truth edit.

\subsubsection{LLM-as-a-Judge.}
Objective evaluations for coding tasks face limitations in our context: direct execution of edited code is infeasible due to the diverse data sources, and we aim to evaluate functionality beyond the three match-based metrics above.
To complement these evaluations, we therefore employ the LLM-as-a-Judge\cite{gu2025surveyllmasajudge} metric, drawing from previous studies\cite{kon2024iac, zheng2023judging}. We selected GPT-4.1 mini, notable for its leading performance as a judge.
The strong correlation observed between these subjective scores and our objective metrics suggests that the LLM judge aligns well with human-like qualitative evaluation.

\subsection{Results}
We provide a detailed analysis based on the evaluation results for all baseline and fine-tuned models, as presented in Table \ref{table:baseline} and Table \ref{table:sft_base}.

\subsubsection{Closed-source models exhibit strong but varied performance, with the Claude and Gemini families leading the field.}
The evaluation results in Table \ref{table:baseline} indicate that Anthropic's and Google's flagship models are top performers. Specifically, Claude 4 Opus and Claude 4 Sonnet demonstrate exceptional capabilities, particularly in Partial Match (67.62\% and 68.57\%, respectively) and Position Match (80.48\% and 80.00\%, respectively). This suggests a superior ability to correctly identify both the content and the location of the required edit. Gemini 2.5 Pro also stands out as a premier model, achieving the highest LLM-as-a-Judge score (68.57\%) among all baseline models, signifying high semantic and logical correctness.

In contrast, the GPT series shows a wider performance variance. While GPT-4.1 is the strongest model in its family, it generally lags behind the top-tier Claude and Gemini models. The GPT-4.1 nano model performs poorly across all metrics, indicating it may lack the necessary capacity for this complex code editing task.

\subsubsection{Open-source models demonstrate highly competitive performance, rivaling many closed-source counterparts.}
The open-source models, DeepSeek-V3 and DeepSeek-R1, deliver impressive results that are comparable to the leading closed-source models. For instance, DeepSeek-R1's LLM-as-a-Judge score of 61.90\% surpasses that of several GPT and Claude variants. The strong performance of DeepSeek models stems from their MoE design, which efficiently allocates resources by dynamically activating only the most relevant expert sub-networks for each input. This specialization enables the model to handle complex tasks more effectively.

\subsubsection{Fine-tuning systematically enhances model performance.}
As shown in Table \ref{table:sft_base}, the Qwen2.5-Coder family exhibits a consistent and positive correlation between model size and performance. The LLM-as-a-Judge score scales smoothly from 42.38\% for the 3B model to 60.95\% for the 32B model. This trend highlights that the \textit{Next Edit Prediction} task significantly benefits from increased model capacity, allowing larger models to better capture the intricate patterns in code context and user intent. The Qwen2.5-Coder-32B model emerges as the top-performing fine-tuned model, achieving scores that are competitive with the elite baseline models.

\subsubsection{Smaller, specialized models can achieve strong performance after fine-tuning.}
The CodeGemma models showcase the effectiveness of fine-tuning on smaller, more specialized architectures. The CodeGemma-7B model achieves a robust LLM-as-a-Judge score of 50.95\%, outperforming even some baseline closed-source models like GPT-4o. This result emphasizes that targeted fine-tuning can make smaller, efficient models powerful contenders for specialized tasks. In contrast, the Codestral-22B model's performance is unexpectedly modest given its size, suggesting that base model architecture and pre-training objectives are crucial for fine-tuning success.

\subsection{Discussion}
\subsubsection{Decoupling \textit{What} from \textit{Where}.}

The relationship between our proposed metrics offers insights. While a high Exact Match score indicates overall proficiency, the discrepancies between Position Match and content-based matches (Partial/Exact) are particularly revealing. For instance, models like Claude 3.5 Haiku exhibit a high Position Match (63.81\%) but a very low Exact Match (17.14\%). This profile suggests a clear decoupling of two distinct underlying skills. The first is Localization (\textit{Where}), the ability to identify the correct location for the next edit, for which a high Position Match score is a strong indicator. The second, distinct skill is Content Generation (\textit{What}), which is the ability to generate the precise code patch for that location. This finding implies that models can be proficient at one aspect while struggling with the other. This diagnostic capability is crucial for future research, as it allows developers to pinpoint and address specific model weaknesses, rather than treating performance as a black box.


\begin{table}[htbp]
\centering
\small
\begin{tabular}{llcc}
\toprule
\textbf{Model} & \textbf{Metric} & \textbf{w/o ft} & \textbf{w/ ft} \\
\midrule
\multirow{4}{*}{\makecell[l]{Qwen2.5-Coder-\\7B-Instruct}}
& Exact Match        & 0.00  & 42.38 \\
& Partial Match      & 19.05 & 46.67 \\
& Position Match     & 34.29 & 65.24 \\
& LLM-as-a-Judge     & 26.67 & 50.00 \\
\midrule
\multirow{4}{*}{codegemma-7b-it}
& Exact Match        & 0.00  & 45.71 \\
& Partial Match      & 6.67  & 48.10 \\
& Position Match     & 13.81 & 68.10 \\
& LLM-as-a-Judge     & 13.81 & 50.95 \\
\bottomrule
\end{tabular}
\caption{Evaluation results of fine-tuned instruct models. “ft” stands for fine-tuning.}
\label{table:sft_instruct}
\end{table}

\subsubsection{Base vs. Instruct.}
To investigate the performance differences between the Base and Instruct models for this task, we conducted tuning and evaluation under identical experimental settings with two models: Qwen2.5-Coder-7B-Instruct and codegemma-7b-it. The results are presented in Table \ref{table:sft_instruct}. By comparing these with the results shown in Table \ref{table:sft_base} for the base version, we observe that, given the same model size, the instruct versions consistently outperform their base version after the same tuning process. We hypothesize that this improvement arises from the instruct tuning stage, during which the model acquires a deeper understanding of the intrinsic relationships between natural language and code editing fragments. This enhanced understanding likely enables the model to better infer user editing intentions in the \textit{Next Edit Prediction} task.

\section{Related Work}

\subsubsection{LLMs for code generation.}  
Code generation has been as a crucial pre-step in automating software development. Early approaches employed encoder-decoder architectures trained on specific code corpora. Notable examples include AlphaCode~\cite{codecontests}, CodeT5~\cite{codet5}, CodeRL~\cite{coderl}, and CodeT5+~\cite{codet5plus}, which demonstrated promising results on various code synthesis tasks. The emergence of decoder-only LLMs has shifted the paradigm, with models like Codex~\cite{codex}, CodeGen~\cite{codegen,codegen2}, InCoder~\cite{incoder}, CodeGeeX~\cite{humanevalx}, StarCoder~\cite{starcoder,starcoder2}, WizardCoder~\cite{wizardcoder}, CodeLlama~\cite{codellama}, MagicCoder~\cite{magiccoder}, DeepSeek-Coder~\cite{deepseekcoder}, and Qwen-Coder~\cite{hui2024qwen2} achieving better performance through auto-regressive pretraining objectives. Meanwhile, general-purpose LLMs trained on mixed data have shown remarkable proficiency in code-related tasks. Models such as Llama3~\cite{llama3}, DeepSeek-R1~\cite{deepseek}, GPT-4~\cite{gpt4}, and Gemini-2.5~\cite{gemini} demonstrate competitive capabilities, often matching or surpassing their code-specialized variants with broader training objectives and larger model size. 

\subsubsection{Code generation benchmarks.}
Various benchmarks evaluate the capability of code generation task of LLMs. Function-level benchmarks include HumanEval~\cite{humaneval}, containing 164 Python problems with function signatures and docstrings, and MBPP~\cite{mbpp}, comprising 974 basic Python problems with functionality descriptions. For comprehensive evaluation, competition-derived benchmarks include APPS~\cite{apps} with 10,000 Python problems of varying difficulty and Code Contests~\cite{codecontests}, a multi-lingual benchmark featuring both correct and incorrect human solutions. Other multi-lingual benchmarks include xCodeEval~\cite{xcodeeval} and HumanEval-X~\cite{humanevalx}. Beyond function and file-level evaluation, research efforts such as RepoEval~\cite{repoeval}, RepoBench~\cite{repobench}, SWE-Bench~\cite{swebench}, and CrossCodeEval~\cite{crosscodeeval} focus on repository-level code generation performance.

\subsubsection{Code editing.}
Code editing tasks focus on modifying existing code to improve functionality, fix bugs, or adapt to new requirements instead of regenerating all the code. Codit~\cite{chakraborty2020codit} trains models using over real-world changes and evaluates them on 5,000 patches to learn specific bug fix patterns from patches. CoditT5~\cite{zhang2022coditt5} introduces a pre-trained model specifically specified for code editing. The study~\cite{cassano2023can} benchmarks the capability of LLMs to follow code editing instructions, highlighting the importance of instruction adherence in automated editing. SarGaM~\cite{liu2024automated} employs searching for related patches, generates code, and modifies it to fit the correct context. Modit~\cite{chakraborty2021multi} presents a multi-modal code editing engine that integrates edit location, code context, and commit messages using a Neural Machine Translator (NMT) framework. jLED~\cite{pian2025you} introduces a novel joint training approach for both localization and source code edits, aiming to provide a more practical solution for real-world code editing scenarios.

\section{Limitations and Future Work}
\label{sec:limitation_future}
While our approach to \textit{Next Edit Prediction} shows promise, it has several limitations that motivate future research. Firstly, our work is constrained by the scale of the available training data. The current training dataset, while adequate for adapting LLMs to the task format, remains insufficient for capturing nuanced patterns of code evolution across diverse editing patterns.
Second, typing our model assumes an idealized top-to-bottom sequential editing pattern, while real-world scenarios often involve non-linear editing behaviors (e.g., revisiting earlier code chunks).
Lastly, the triggering condition for predicting the next edit relies on implicit user-IDE interactions (e.g., typing pauses, newlines). This assumption may not always hold, as users might not intend to make a related subsequent edit at every such juncture.

Looking ahead, we identify two primary directions for future work.
First, we aim to develop refusal mechanisms that enable models to abstain from predictions when confidence is low, thereby minimizing disruptive suggestions.
Furthermore, we plan to enhance the contextual understanding of our model by incorporating cross-file context, better aligning with developers' holistic workflow.

\section{Conclusion}
Accurately predicting the next code edit by inferring user intent, as pursued in the \textit{Next Edit Prediction} task, represents a critical advancement in the development of AI-assisted programming. This capability signifies a shift from reactive code completion tools toward proactive AI systems that can comprehend developer goals, anticipate forthcoming actions, and potentially automate more sophisticated tasks such as code refactoring or feature implementation. The strong empirical performance, particularly in generating contextually appropriate and largely accurate suggestions, indicates that the vision of deeply integrated, intelligent AI collaborators in software engineering is becoming increasingly attainable.

\bibliography{ref}

\end{document}